\RequirePackage{fix-cm}
\documentclass[smallextended]{svjour3}       %
\smartqed  %
\usepackage{graphicx}
\usepackage[hidelinks,backref=page,colorlinks=false,pdfborder={0 0 0},hyperfootnotes=false,breaklinks=true]{hyperref} 

\usepackage{amsmath}
\begin{document}

\title{The shortness of human life constitutes its limit%
}

\author{Brandon Milholland         \and
        Xiao Dong \and
Jan Vijg%
}

\institute{B. Milholland\at
  Department of Genetics \\
Albert Einstein College of Medicine \\
              1300 Morris Park Avenue \\
              Bronx, NY 10461 \\
              USA \\
              Tel.: +1 (718)-430-2000\\
              \email{brandon.milholland@phd.einstein.yu.edu}           %
}

\date{Received: date / Accepted: date}

\maketitle

\begin{abstract}
In this paper, we affirm our earlier findings of evidence for a limit
to human lifespan. In particular, we assess the analyses in extreme
value theory (EVT) performed by
Rootz\'en and Zholud. We find that their criticisms of our work are
unfounded and that their analyses are contradicted by several other
papers using EVT. Furthermore, we find that even if we completely accept the
conclusions about late-life human mortality reached by Rootz\'en and
Zholud, their results do not actually contradict the findings
presented in our original paper: whether unbounded or not, human
lifespan is unlikely to greatly exceed 120 years, and the
improbability of longer survival---whether it is exactly zero or
merely astronomically small---acts as a \textit{de facto}
limit. In order to eliminate the confusion surrounding the issue, we
propose the adoption of the term ``limit'' to denote the age at which
the chance of survival is exactly zero and the term ``effective
limit'' to denote the age at which the change of survival falls below
a given threshold. Once this distinction is made, it can be
demonstrated that the final result of Rootz\'en and Zholud is
essentially a recapitulation of the main conclusion of our paper.  Ultimately, much of the controversy
 surrounding the issue of a limit to human lifespan can be avoided by
 carefully reading the literature and applying statistics to practical
 human scales.

\end{abstract}

\section{Introduction}
\label{intro}
In their paper \cite{rootzen_human_2017}, Rootz\'en and Zholud assert
that there is no limit to human lifespan and also level several
criticisms at our paper \cite{dong_evidence_2016} in which we reported evidence
for such a limit. Harsh words, such as ``unfounded'',
``inappropriate'', ``wrong'' and ``misleading'' are used, but the
severity of their language belies the fact that their paper is the one
that is misleading by misrepresenting our findings and attacking us on
tenuous, incorrect and unfounded bases (Section \ref{sec:refuting}). A close examination of the work of
Rootz\'en and Zholud will reveal that their analysis is inadequate to
conclusively demonstrate that human lifespan is
unlimited (Subsection \ref{subsec:probs}). Furthermore, several other analyses have come to
conclusions the opposite of Rootz\'en and Zholud (Subsection \ref{subsec:contradictions}). Finally, even Rootz\'en
and Zholud's own work suggests that there is a \textit{de facto} limit
to human lifespan, one that is not much higher than the one we found
in our original paper (Section \ref{sec:shortness}). In one very narrow sense---one that we did not
propose at all in our paper---human lifespan may be
unlimited. But, in a much more practical and reasonable sense, human
lifespan, given the current paradigm of medical research, is limited. 

\section{Refuting the criticisms of Rootz\'en and Zholud}
\label{sec:refuting}
In Section 4 of their paper, Rootz\'en and Zholud summarily dismiss the
the results presented in our paper poiting to a limit to
human lifespan \cite{dong_evidence_2016} in four brief
paragraphs\footnote{Rootz\'en and Zholud also cursorily mention  our findings that
  improvements in mortality decrease with age and argue that `` a slower rate of improvement is still an improvement,
and, if anything, it contradicts the existence of a limit'', but this
assertion is never backed up by any evidence or reasoning. A slower
rate of increase does not imply that the increase will continue
forever, nor does it imply that there is no limit. By Rootz\'en and
Zholud's logic, an examination of some partial sums of the series
$\sum_{i=1}^{n} 2^{-i} $ would show a slowing rate of
increase as $n$ increases, which ``if anything \ldots{} contradicts the existence of a limit''. However, it has been mathematically proven that geometric
series in this form converge (i.e., $\lim_{n\to\infty} \sum_{i=1}^{n}
2^{-i} =1$), so there is in fact a finite limit to the sum.
Furthermore,  Rootz\'en and Zholud do not address the fact that that there has been not merely a deceleration of
improvement but an absence of any detectable improvement whatsoever in
supercentenarian mortality.}. These paragraphs only
address Figure 2 of our paper and do not address the findings
presented in Figure 1 and Extended Data Figures 1, 2, 3, 4, 5 and 6 of
our paper. In these four crucial paragraphs,
Rootz\'en and Zholud begin with a falsehood, move on to a potential
self-contradiction, make a nonsensical statement, and finally conclude
by fundamentally misunderstanding the thesis of our paper.

In the first paragraph, Rootz\'en and Zholud state that the number of
supercentenarian deaths in the International Database of Longevity (IDL) varies from 0 to 42. However, this
statement is false: it is plain to see from Rootz\'en and Zholud's
Figure 6 and Figure 7 that the minimum number of deaths in any year is
1. 

Rootz\'en and Zholud also assert in the first paragraph of Section 4 that the yearly
maximum reported age at death (MRAD) ``shows the same pattern'' as the
number of deaths; unfortunately, they do not provide any correlation
statistics that could be used to evaluate that assertion. In the
next paragraph, they mention a different line (representing $110+$
mean of maximum $n_t$
exponential variables) that they also assert, without statistical substantiation, ``agree[s] well'' with
the MRAD, but this line differs greatly from the line representing the
number of deaths (most visible in the interval 1990--2000). It would
seem contradictory that the line representing number of deaths, which
fluctuates in magnitude several-fold over short periods of time (for
example, leaping from 5 to 35 in the early 1990s), and the line representing $110+$
mean of maximum $n_t$
exponential variables, which is devoid of such wild fluctuations
(barely ambling from around 113 to around 115 over the same interval), could both
strongly correlate with the MRAD. Perhaps the graphs are misleading,
and both variables, counterintuitively, manage to correlate well with
the MRAD, but Rootz\'en and Zholud provide no statistical evidence that
they do.

In the third paragraph, Rootz\'en and Zholud say that our findings
result from ``inappropriate combination of data from different time
periods''. This does not make sense: we merely used all the data from the
entire time interval of the IDL (and, separately, the GRG). Presumably, they meant to say that
there was an inappropriate combination of data from different
\textit{countries} since that is the issue they touch upon at the end
of the previous paragraph. However, considering each country
separately still finds evidence for a trend break, contrary to Rootz\'en and Zholud's
assertion. In Japan, all of the data is post-breakpoint, so
there is no evidence for or against, although the number of
supercentenarian deaths does increase more rapidly than the MRAD,
suggesting the two have become decoupled. In the USA, the increase of the
MRAD continues unabated, providing some evidence against a trend
break; but note also the sharp jump in the number of supercentenarian
deaths in the early 1990s and the comparatively shallower increase in
MRAD, again undermining Rootz\'en and Zholud's thesis that the MRAD is driven
by number of supercentenarian deaths. In England and Wales, the slope
of the trend decreases from 0.7 to 0.2. Rootz\'en and Zholud do not
define what evidence is sufficient to indicate a trend break,
prompting the question: if a more than 3-fold decrease
in the slope is not evidence for a trend break, then what is? Finally,
in France the slope goes from 0.2 to --0.3, indisputable evidence of a
trend break. To summarize, of the four countries considered, one of
them is invalid, one arguably does not have a trend break, and two
almost certainly do. Furthermore, all four contain fluctuations in the
number of supercentenarian deaths not reflected in the
MRAD. Therefore, the balance of evidence favors the conclusions
presented in our original paper.

The final paragraph dismisses the evidence of a limit in the 2nd
through 5th highest reported ages at death with a repetition of the
nonsensical accusation of ``inappropriate combination of data from different time
periods'' and no substantiating evidence. Given the weakness of the
reasoning employed by Rootz\'en and Zholud regarding the MRAD, it is
likely that this assertion regarding the 2nd through 5th highest
deaths is not supported by the evidence, but their thesis is ultimately
impossible to evaluate due to the absence of information provided in
the paper. Section 4 of Rootz\'en and Zholud's paper terminates with a
fundamental misinterpretation of the thesis of our paper. The main
observation of our paper was that the MRAD had
stagnated since the mid-1990s and, despite advances in medicine,
fluctuated around 115, without further increase; this stagnation
around the average of 115 constitutes the ``limit'' referred to in our
paper. As a corollary, if one desires to know the absolute highest age to
which any human could reasonably expect to live, then we also
calculated that an MRAD of more than 125 would be expected only once
every 10,000 years; this value may also be considered a ``limit'' to
human longevity. In short, the term ``limit'' is overloaded, and can
refer to distinct, yet related, concepts depending on context. In
their paper, Rootz\'en and Zholud introduce a third definition of
``limit'': the point at which the chance of survival is equal to
zero. They contend that since the distribution of MRAD ages is
unbounded, there is no ``limit'' to human lifespan. On its own, there
is nothing fundamentally wrong with this contention, but the failure
of our paper to conform to the arbitrarily altered definition of
``limit'' formulated by Rootz\'en and Zholud cannot be considered an
error on our part. Indeed, as we will see, the definition of ``limit''
used by Rootz\'en and Zholud is not very useful, as it would apply even
to situations in which the probability of survival is so small as to
be negligible. From a very philosophical perspective, it is possible
to argue that due to the inherently stochastic nature of the universe
implied by quantum physics, it is impossible to ever assign a
probability of zero to any event, so the conclusion reached by Rootz\'en
and Zholud is not novel. Rootz\'en and Zholud build their paper on a
more immediately applicable starting point---demographic data---but if their goal
was to derive the conclusion of no limit to lifespan based on that
foundation alone, they have made several omissions and questionable
decisions that undermine the strength of their analyses.

\section{Problems with the analyses of Rootz\'en and Zholud and
  literature contradicting their results}
\label{sec:problems}

Having made our reply to the criticisms Rootz\'en and Zholud leveled
against our paper, we move on to several of our own criticisms of
their paper. The conclusion reached by Rootz\'en and Zholud---that human
life is ``unlimited''---is extraordinary, even if qualified by ``but
short''. Such a strong conclusion requires strong evidence, but that
which is
presented by Rootz\'en and Zholud is not up to the task. Furthermore,
the conclusions reached by Rootz\'en and Zholud put them at odds with
multiple other papers, not just our own. These contradictions must be
resolved before the contention of an unlimited human lifespan can be accepted.

\subsection{Problems with the analyses of Rootz\'en and Zholud}
\label{subsec:probs}

Much of the work in Rootz\'en and Zholud's paper is of good quality, but
there are several areas where they use statistics inappropriately or
come to conclusions that do not necessarily follow from the premises
given. These deficiencies do not discredit the entire paper, but they
significantly weaken the strength of the findings presented
therein. Indeed, it is possible that each could be addressed and a
confirmation of Rootz\'en and Zholud's analysis found. Therefore, we
present these issues in the spirit of a scientific dialogue, in the
hopes that further investigation will advance our knowledge, either
for or against Rootz\'en and Zholud's conclusions.

First, the data used by Rootz\'en and Zholud appear to support a
stagnation of supercentenarian mortality, with no improvement in
supercentenarian survival in recent decades, something that would be
consistent with a limit to human lifespan and not a continual
improvement towards breaking that limit.
In Table 3 of their paper, Rootz\'en and Zholud present the results of
statistical tests in several sets of data for differences in
supercentenarian mortality
between the first half of each set (generally spanning from the 1960s
until the 1990s) and the second half (generally from the mid-to-late
1990s until the mid-to-late 2000s). In every single case, they find no
evidence of a change in mortality, with the lowest $p$-value being
0.18, not even close to significant. This finding essentially confirms
our earlier results\cite{dong_evidence_2016}, i.e. a lack
of improvement in supercentenarian mortality indicates that human
lifespan has stopped increasing, and is essentially a restatement of
the finding presented in Figure 3c of our original paper.

Second, Rootz\'en and Zholud have chosen an inappropriate null
hypothesis and incorrectly interpreted the results of their
statistical tests.
Rootz\'en and Zholud begin with the assumption that the force of
mortality after age 110 is constant, and that human lifespan therefore
follows an exponential distribution. Then, they provide several
$p$-values, in Table 5 of their paper, which fail to reject this null
hypothesis. However, it is basic knowledge regarding $p$-values that a failure to reject the null hypothesis does not
mean that the null hypothesis is true. Furthermore, the assumption of a constant force of mortality is not an
appropriate null hypothesis. From early adulthood onwards,
mortality increases relentlessly; it is  far more parsimonious
to assume that this increase will continue rather than that
it will stop in its tracks at old age. Indeed, by making a constant
force of mortality (which corresponds to an ``unlimited but short''
lifespan) their null hypothesis, Rootz\'en and Zholud have
committed the fallacy of assuming that which was to be
demonstrated. Finally, despite having tilted the scales in favor of a
constant force of mortality, Table 2 of Rootz\'en and Zholud's paper
indicates that the GRG dataset, the data with the greatest temporal
and geographic reach, strongly rejects the hypothesis of an
exponential distribution.  Thus, the main premise of Rootz\'en and
Zholud's paper, that the force of mortality after age 110 is constant,
is not supported by the data and statistics presented.

\subsection{Contradictions with the literature}
\label{subsec:contradictions}

Having established that Rootz\'en and Zholud failed to demonstrate a
constant force of mortality, we should not be too harsh. After all,
the task they have set out for themselves is impossible: showing that
the force of mortality is constant would require showing that there is
no change in mortality, which would require proving a negative. A more
realistic goal would be to calculate a confidence interval for
the force of mortality after age 110 in order to establish an upper
bound for mortality. That has been done by others\cite{modig_how_2017}, and the
upper bound of the
resulting interval rapidly approached 1. Although the confidence
intervals were wide, leading the authors of that paper to conclude
that mortality after age 100 plateaued, the fact that
the estimate of mortality also rapidly increased past 0.5 after age 110 and rapidly
approached 1 indicates that the balance of evidence is actually in favor of a
limited lifespan as defined by Rootz\'en and Zholud. Furthermore, they
found that centenarian mortality has not improved, a result that
strongly supports our findings of a limited human lifespan based on MRAD.  As the authors of
that paper concluded: ``the maximum lifespan, measured as the age of
the oldest person to die, is currently not increasing''.

Rootz\'en and Zholud have used EVT in their attempt to resolve the
question of whether human lifespan is limited, reaching the
conclusion that it is unlimited. Three other recent papers have
examined the same question using EVT and come to the opposite
conclusion. The first\cite{gbari_extreme_2017} examined data from 46,666
Belgians who died at 95 or older and found an ``ultimate age'' of 115
for men and 123 for women, a result very similar to our own.  The
second paper\cite{feifel_who_2017} analyzed data from the International Database on
Longevity (IDL) and the Human Mortality Database (HMD). The authors of
that paper write that ``due to its small sample size, the hypothesis of an infinite lifespan could
not be rejected for the IDL,'' the dataset used by Rootz\'en and
Zholud, but when they supplemented it with data from the HMD, ``we found significant evidence for a finite
lifespan in the combined data set and obtained reliable point
estimates for the maximum attainable age'' of 125--128 years in
females. Finally, a paper\cite{einmahl_limits_2017} based on precise measurements of the ages at
death of 285,000 residents of the Netherlands found ``compelling
statistical evidence'' for a finite human lifespan in both men and
women. They found an average annual endpoint to the lifespan
distribution of 114 and 116 years and a
maximum estimated upper endpoint of 125 and 124 years in men and women
respectively\footnote{The authors of that paper also found a lack of a
  trend in with time in these values; although they imply that the
  lack of a trend would contradict our paper, this implication is
  based on a misinterpretation of our results, which found an initial
  upward trend, followed by a lack of a trend (and not necessarily a
  decrease, as indicated by the insignificant $p$-value) during recent
  decades. Instead, the lack of a trend is a confirmation of our findings
that there has essentially been a limit in place the whole time, which
was finally reached in the mid-1990s. }. These estimates are very close to the estimates of
average MRAD, 115 years, and maximum MRAD, 125 years, which we arrived
at in our original paper. To summarize, all three of these other EVT
papers confirm our results and contradict those of Rootz\'en and Zholud.

To conclude this section, we would like to use Rootz\'en and
Zholud's concluding sentences to tie together the two
previous strands: the gaps in their paper and its variance with the
literature. In the penultimate and final sentences of their paper,
Rootz\'en and Zholud concede that ``the
IDL data set only includes 9 persons who lived longer than 115 years, so data is sparse
above this age'' but assert that ``these 9 data points agree well with
our conclusion''. No substantiation is given for this assertion, and
the very sparsity of such data is itself suggestive of a limit to
human lifespan, but
even if it were true that those 9 data points did support Rootz\'en and
Zholud's conclusion, they must surely be outweighed by the
contradictory studies based on over 300,000 data points.

\section{Shortness of life as a \textit{de facto} limit: theoretical
  considerations and empirical results }
\label{sec:shortness}

\begin{figure*}
  \includegraphics[width=0.75\textwidth]{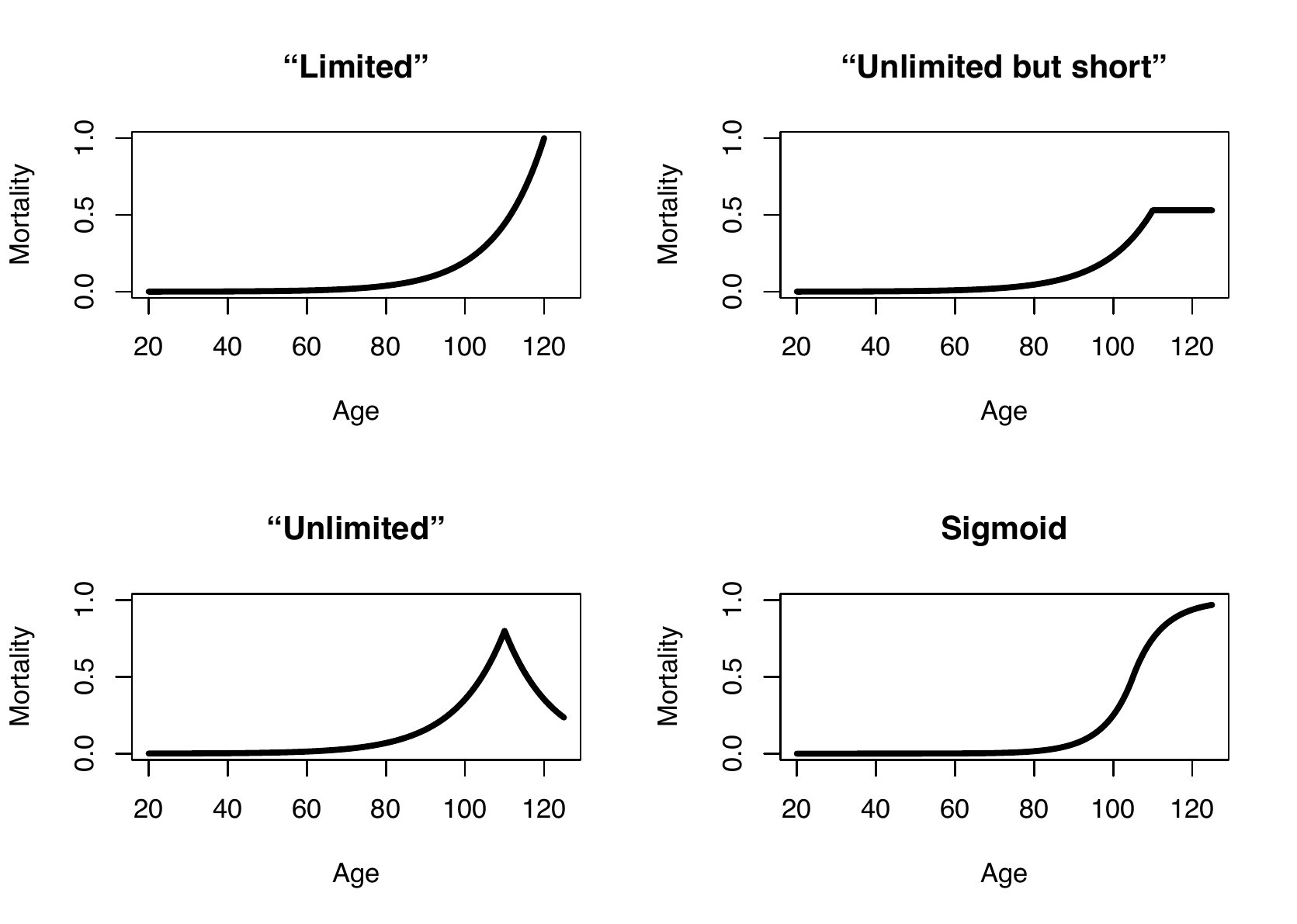}
\caption{The three trajectories of late-life mortality envisioned by Rootz\'en and Zholud, and one
they do not consider, placed in context of the earlier exponential
increase in mortality. In the top left corner, the ``limited''
scenario, mortality increases exponentially throughout life until it
hits 1 (certain death). In the top right corner, the ``unlimited but
short'' scenario, mortality increases exponentially until it reaches
0.53 and then remains stationary. In the bottom left corner, the
``unlimited'' scenario, mortality increases exponentially and then
decreases late in life. Finally, in the bottom right corner, the
sigmoid scenario not considered by Rootz\'en and Zholud, mortality
follows an approximately exponential increase early in life until it
begins to level off and asymptotically approach 1; mortality never reaches 1, but it does not
decline or stagnate and continues to increase at an ever-slower
rate. Note the lack of parsimony in both of the ``unlimited''
scenarios proposed by Rootz\'en and Zholud. }
\label{fig:trajectories}       %
\end{figure*}

As we have shown, there is considerable evidence against Rootz\'en and
Zholud's conclusion that the force of mortality remains constant after
age 110. However, there still remains a slim possibility that they are
correct. It is also possible that the force of mortality could
increase with age and asymptotically approach 1 without ever reaching
it, such as in a sigmoid function (Figure \ref{fig:trajectories}), a scenario not considered
by Rootz\'en and Zholud. In any case, let us assume that Rootz\'en and Zholud are correct
there is no limit, $L$, beyond which the probability
of survival is 0. Would this contradict our results? And would it be a
finding of importance? The answer to both questions is: no.

First, a hard limit, with a chance of mortality equal to 1, was never part of our
original paper. As a commentary on our paper
observed\cite{olshansky_ageing:_2016}, our paper's thesis is that ``there is no fixed limit
beyond which humans cannot live, but that there are, nevertheless,
limits on the duration of life''; the ``limit'' referred not to an age
of certain death but the stagnation of advances in human maximum
lifespan and the corollary emergence of an age beyond which survival,
although not impossible, is prohibitively unlikely. To refer directly
to the text of our paper: ``we found that the probability of an MRAD
exceeding 125 in any given year is less than 1 in 10,000''---not
zero. In light of this, Rootz\'en and Zholud's statement that ``it is
likely that the record age \ldots will be shorter than 128 years''
is neither contradictory to our conclusions nor particularly novel,
but rather a slightly different estimate of the number: stated
informally (we shall formalize it below), the age which MRAD is
unlikely to exceed in the foreseeable future.

The main thesis of Rootz\'en and Zholud's paper, that there is no
``limit'' by their definition to human lifespan is not particularly
meaningful or novel. Indeed, it is not even necessary to consult any demographic data in order to
arrive at the conclusion that there is
no age with a zero chance of survival. According to quantum physics,
anything is possible, even if extremely unlikely. So, the odds of a
brain spontaneously appearing out of thin air due to quantum
fluctuations have been calculated\cite{linde_sinks_2007} to be $1$ in $10^{10^{50}}$. Since aging is due to physical
changes, i.e. the accumulation, depletion or modification of
macromolecules\cite{lopez-otin_hallmarks_2013}, it is possible that quantum
fluctuations could spontaneously transform a supercentenarian's body
to that of a 20 year old, making them a shoo-in for a new longevity
record. Calculating the exact probability of this event occurring, and the question
of whether the number of zeros in its decimal representation would
exceed the number of atoms in the observable universe, we leave as an
exercise for future investigators.

In case the scenario described above strikes readers as too
outlandish, we propose a more realistic occurrence based on Rootz\'en and
Zholud's model of ``unlimited'' lifespan: assuming that Rootz\'en and
Zholud are correct and mortality after age 110 follows the ``each year
a coin is tossed'' model, and rounding\footnote{Observant readers will
  note that the abstract of  Rootz\'en and
Zholud's paper is contradicted by its body: in the former, the
probability of dying is given as 47\%, while in the latter that value
is given as the probability of survival. While mixing up life and
death is a major error, the proximity of both probabilities to 50\%
limits its effects. }  up the probability of survival
to 50\%, the odds of a supercentenarian living to age 150 are $1$ in
$2^{40}$ or just under 1 in a trillion. So, on the one hand we are
told that ``human life is unlimited'' and on the other hand the same
model gives probabilities of survival that are so low that many would
consider them next to impossible. At a certain point, discussing the
probability of a single individual living to a given age under Rootz\'en and Zholud's
``unlimited'' lifespan paradigm becomes akin to discussing the
probability of receiving a single molecule of the active ingredient
from a homeopathic remedy.

Our point is that the finding that human lifespan is ``unlimited'' on the
basis that there is no age with a zero chance of survival may be
mathematically true and it may be of theoretical or even recreational
interest, but it is of little practical significance or use. Insisting
that a ``limit'' to lifespan consist of an age with 100\% mortality
gives the term ``limit'' too narrow and restrictive a definition, one
we did not use in our original paper and one
that makes any statements regarding its existence or nonexistence
almost completely inconsequential. Instead of discussing the limit,
$L$, of lifespan as the age with a zero chance of survival, we propose
that future discussions of extreme longevity focus on the effective
limit, $L_e$ of lifespan, the age beyond which it is extremely
unlikely any individual of the species will survive. The value of
$L_e$ will depend on many factors, the most pressing of which is the
definition of ``extremely unlikely''. Therefore, $L_e$ can be stated
formally by qualifying it with $\epsilon$, the desired definition of
``extremely unlikely''. For example, our original paper calculated
$L_e^{\epsilon=0.0001}=125$, i.e. the probability of any human surviving past
125 is 0.0001 or 1 in 10,000. In their paper, Rootz\'en and Zholud essentially
calculate $L_e$ to be 128, but do not provide the value of $\epsilon$
they used. The values of $L_e$ and $\epsilon$ can be inferred from
other publications (Table \ref{tab:Le}). Generally,
$L_e^{\epsilon=0.5}$ is around 115, while values of $L_e$ for lower
values of $\epsilon$ are in the mid-120s. 

\begin{table}
\caption{Values of $L_e$ and $\epsilon$ inferred from other
  publications. }
\label{tab:Le}       %
\begin{tabular}{llll}
\hline\noalign{\smallskip}
$L_e$ & $\epsilon$ & Data source & Citation  \\
\noalign{\smallskip}\hline\noalign{\smallskip}
114.9 & 0.5 & IDL/GRG & \cite{dong_evidence_2016} \\
125 & 0.0001 & IDL/GRG & \cite{dong_evidence_2016} \\
131.21 & $\ll{}0.025$ & Belgian National Population Register &
\cite{gbari_extreme_2017} \\
128.73 & $\ll{}0.025$ & IDL/HMD & \cite{feifel_who_2017} \\
115.7 & 0.5 & Statistics Netherlands & \cite{einmahl_limits_2017} \\
123.7 & 0.025 & Statistics Netherlands & \cite{einmahl_limits_2017}  \\
115 & 0.5 & ``Best-guess'' MRAD & \cite{milholland_best-guess_2017} \\
128 & ? & IDL/GRG & \cite{rootzen_human_2017} \\
\noalign{\smallskip}\hline
\end{tabular}
\end{table}

The value of $L_e$, for a given value of $\epsilon$, is not
constant. It is determined by two factors: the population of the
species (more individuals means more chances for one of them to reach
an extreme age) and the chance of survival at each age leading up to
$L_e$. In our original paper, we came to the conclusion that $L_e$ had
basically reached \textit{its} limit---here, meaning that $L_e$ had
remained for years and would remain in the absence of unprecedented
technological breakthroughs at its current level---based on the tacit assumption
that the human population would not greatly increase and our explicit
demonstration that late-life survival had stagnated. Rootz\'en and
Zholud speculate that future research may further extend human
lifespan, and we do the same in the conclusion of our original paper,
stating that ``there is no scientific reason why such efforts could
not be successful''. However, the gains from such efforts have not
been realized. There is currently no cure for aging, nor is there even
a product approved to be marketed as a treatment for aging. Perhaps
one day human lifespan will become unlimited, but right now the
chances of mortality at old age act as an effective limit to lifespan.

\section{Conclusion}
\label{sec:conclusion}

The premise and even the title of Rootz\'en and Zholud's paper are
self-contradictory: ``human life is unlimited --- but short''. We
resolve this contradiction by positing that the shortness of human
life constitutes its limit. For a particular, narrow definition of
``limit'' it can indeed be said that there is no limit to human
lifespan, but for a more useful definition there is one. If one finds it justified to say that human life is ``unlimited''
because there is a constant 47\% chance of survival each year, one
must also find it justified to say that roulette winnings are
``unlimited'' because there is a constant 48.7\% chance of doubling
one's money by betting on black every time. It is easy to see the
problem with relying on the latter reasoning for a moneymaking
strategy, so why would anyone accept the former reasoning for a
longevity strategy?

The use of an overly restrictive definition of ``limit'' (as detailed
in Section \ref{sec:shortness}) is not the
only issue with Rootz\'en and Zholud's paper. They also make unfounded
criticisms of our findings of evidence for a limit to human lifespan
(refuted in Section \ref{sec:refuting}) and the evidence even for
their modest conclusion is weak, with both statistical issues
undermining their findings (enumerated in Subsection
\ref{subsec:probs}) and larger studies having contradictory results
(reviewed in Subsection \ref{subsec:contradictions}). Thus, the
contention that ``human life is unlimited'' is most likely to be
false. 

 Much of the controversy surrounding this issue could have been
 avoided by carefully reading our paper and using the context of the
 word ``limit'' as we used it to correctly infer its meaning in that
 paper. Imposing an inappropriate definition of the word does not
 contribute to the literature, as it seeks to rebut our paper for a thesis it
 does not actually advance. 

However, there is some value to Rootz\'en and Zholud's work, as it has
drawn attention to the necessity of explicitly formulating a definition of
``limited'' that is useful for discussions of human lifespan. Although
the evidence available suggests that, contrary to Rootz\'en and Zholud's
conclusion, the risk of mortality does not remain constant after age
110, there also is no age at which mortality is certain. In that
sense, then, human life is unlimited: for an infinitely large cohort
or for an infinitely many cohorts, there is no maximum age beyond
which no individual will live. However, for finitely-sized cohorts,
observed over a sensible interval of time (10,000 years, being a few
times longer than the duration of recorded civilization, is probably
the absolute maximum that can be seriously considered) the age-related
elevation of mortality acts to constrain the highest age that is
likely to be observed, with estimates of that constraint clustering
around 125. In order to distinguish between these two concepts, we proposed
the term ``effective limit'' to indicate the age at which the
probability of a single individual surviving is negligible (falling
below a given threshold). We hope that this new term will be useful in
facilitating clear and meaningful discussion of longevity in the
future.

\bibliographystyle{spmpsci}      %
\bibliography{Rootzen_reply}   %

\end{document}